**Article type**

Full Length Article

**Research on the Influence Mechanism and Effect of Digital Village Construction on Urban-Rural Common prosperity——Evidence From China**

Huang Dahu[a], Shan Tiecheng[b*], Wang Cheng[c],

[abc] College of Economics & Management, Hubei University of Arts and Science, Xiangyang, 441053, China

* Corresponding author:15527180307@163.com(Shan Tiecheng)

**Abstract**

Urban-rural common prosperity is the ultimate goal of narrowing the gap between urban and rural areas and promoting urban-rural integration development, and it is an indispensable and important element in the common wealth goal of Chinese-style modernization. Based on the panel data of 30 provinces (autonomous regions and municipalities) in mainland China, the impact effect and mechanism of digital village construction on urban-rural common wealth are empirically examined by constructing the mediation effect model and spatial econometric model for the period of 2011-2022. The study finds that： firstly, digital village construction can significantly promote urban-rural common prosperity, and the conclusion remains robust after replacing the explanatory variables and the measurement model. Secondly, the promotion of agricultural technology progress and rural labor force transfer is an important path for digital rural construction to promote the realization of urban-rural common prosperity. Finally, there is an obvious spatial spillover effect of digital village construction in promoting urban-rural common prosperity, in other words, digital village construction can not only effectively promote urban-rural common prosperity in the region, but also have a positive effect on the development of urban-rural common prosperity in neighboring regions. Based on this, it is proposed that a differentiated and dynamic development strategy be adopted to accelerate digital construction in rural areas, strengthen interregional exchanges and cooperation, and optimize the spatial layout of digital rural development.

**Keywords**

## 1.Introduction

Common prosperity is an essential requirement of socialism with Chinese characteristics and an important feature of Chinese-style modernization. Since the 18th National Congress of the CPC, China has orderly pushed forward the citizenship of the agricultural transfer population, continued to promote the reform of the rural land system, accelerated the construction of infrastructure in urban and rural areas as a whole, coordinated and pushed forward the equalization of basic public services in urban and rural areas, and won the battle against poverty through various important initiatives, making significant progress in overcoming the problem of unbalanced urban-rural development and breaking up the urban-rural dichotomy, and laying a solid foundation for the promotion of common wealth in both urban and rural areas. A solid foundation has been laid for promoting common prosperity in urban and rural areas. However, among the three major gaps, namely regional, urban-rural and group, the urban-rural multidimensional gap is still a serious constraint to the substantive advancement of common prosperity in China (Kong Xiangzhi et al., 2022; Xi Heng, 2023). According to the National Bureau of Statistics, the ratio of disposable income per capita between urban and rural residents in China was 2.45 in 2022, which is significantly higher than that of developed countries such as the United States (1.38), Japan (1.14), and South Korea (1.43), as well as developing countries such as India (1.44) and Vietnam (1.19) (Guo Yan et al., 2022). The report of the 20th Party Congress points out that "the most arduous and burdensome task in building a socialist modernized country in an all-round way still lies in the countryside". Correctly handling the relationship between urban and rural areas is not only an objective need to achieve the second hundred-year goal, but also a major issue that concerns the overall situation of Chinese-style modernization. Therefore, it is necessary to place towns and villages on an equal footing in terms of complementarity and interaction, and to make every effort to promote the common wealth of urban and rural areas in terms of production and management, public services and the ecological environment, so as to ensure that urban and rural residents share the fruits of modernization and development.

Urban-rural common wealth is the ultimate goal of narrowing the gap between urban and rural areas and integrating urban and rural areas, and it is an important and indispensable part of the goal of modernization and common wealth in Chinese style. Scholars have carried out useful explorations focusing on the connotation, level measurement and influencing factors of urban-rural common wealth. From the viewpoint of connotation, urban-rural common wealth is the organic unity of active government and effective market complementing and promoting each other, ultimately realizing "reciprocity between workers and peasants" and coordinated development of urban and rural areas (Sun Fang, 2023), and including urban-rural universal wealth, urban-rural common wealth, urban-rural comprehensive wealth and urban-rural gradual wealth (Zhang Minghao et al., 2023), as well as the organic unity of urban-rural universal wealth, urban-rural common wealth, urban-rural comprehensive wealth and urban-rural gradual wealth (Zhang Minghao et al., 2023). In terms of measurement, some scholars have adopted the urban-rural income gap to characterize urban-rural common wealth (Keqi Ma, 2024), and more scholars have measured the level of urban-rural common wealth by constructing a comprehensive indicator system (Lin, 2024; Yang Heping, 2024). In terms of influencing factors, digital finance (Yang Fang, 2023), urban meandering production (He Yaxing, 2023), digital innovation and entrepreneurship (He Lu, 2024), and digital government construction (Xin Lulu, 2024) are all able to reduce the urban-rural development gap and promote urban-rural common wealth to different degrees.

Since the 21st century, modern information technology innovations such as big data, the Internet of Things, and artificial intelligence have flourished, giving rise to a series of new products, technologies, and business models, and promoting deep changes in industrial forms and economic development patterns. Big data and other modern information technologies have unique advantages in smoothing the flow of urban and rural factors (Hu Yonghao et al., 2024), promoting the integration of rural industries (Wang Dingxiang et al., 2022; Yuan Xiaohui et al., 2024), and strengthening rural social governance (Shen Feiwei, 2020), which provide unprecedented new opportunities for promoting the integrated development of urban and rural areas and realizing the common wealth (Zeng Yiwu et al., 2021; Chen Jiemei et al. 2023). Some studies have shown that digital rural construction plays an important role in promoting the growth of farmers' income and consumption (Wang Yannan, 2021; Shi Changliang, 2023; Ding Jianjun et al., 2023), the high-quality development of agriculture (Xia Xianli, 2019), and the development of new industries and new forms of business in rural areas (Yin Haodong, 2020). In fact, digital technology-enabled rural construction has become an effective hand in narrowing the gap between urban and rural areas and moving towards the goal of common wealth (Chen Guisheng, 2022).

The above studies have laid a good foundation for the holistic study of urban-rural common wealth. However, from the perspective of existing studies, there are still some points that need to be supplemented: firstly, existing studies have explored the impact of digital villages on common wealth, but rarely focused on urban-rural common wealth; secondly, there is still a lack of exploration of the mechanism of digital village construction to promote the realization of urban-rural common wealth. In view of this, this paper tries to construct a multi-dimensional indicator system to measure the development level of digital village construction and urban-rural common wealth, in order to analyze the impact of digital village construction on urban-rural common wealth, and tries to reveal the mechanism of the impact of digital village construction on urban-rural common wealth from the paths of technological progress and labor force transfer, with a view to providing a reference for the implementation of the policy of promoting the construction of China's digital villages under the goal of common wealth.

## 2.Theoretical analysis and research hypotheses
*2.1.Direct effects*

With the development of big data, artificial intelligence and other technologies, digital information technology sinks into rural areas and integrates with the local economy and society, which can revitalize all kinds of production factors in the countryside, enhance the flow of information and resources between the countryside and the outside world (Liu Shaojie et al., 2021), and endow the countryside with new kinetic energy for development. For a long time, the direct obstacle that restricts the reduction of the urban-rural income gap lies in the slow growth of rural residents' income (Li et al., 2019), and sustained improvement of rural residents' income is the key to narrowing the urban-rural income gap and realizing the common wealth of urban and rural areas. Digital information technology has gradually shown positive effects in increasing farmers' income, improving farmers' quality of life, and improving rural social governance (Xie Lu et al., 2022). First of all, the construction of digital countryside can improve farmers' income and boost the common prosperity of urban and rural areas. The development of digital technology can make direct matching between buyers and sellers without the need to go through intermediaries (Bakos, 1998), which fills the short board of rural residents in market information acquisition and sales channels (Shimamoto, 2025), thus allowing rural residents to obtain more profits and increase productive income. At the same time, the development of digital technology has spawned new businesses and so on, which provide more several pages of opportunities for farmers and help increase their wage income. Secondly, digital village construction can improve rural social governance and help promote common prosperity in urban and rural areas. Digital technology provides more effective and convenient tools for social governance in rural areas, which is conducive to improving the efficiency of social governance and narrowing the gap between urban and rural social governance (Zhao Xiuling, 2019). For example, the information platform constructed by mobile Internet and other technologies embedded in rural governance can realize interaction over time and space and participation at low cost (Qiu Zeqi et al., 2022), which can strengthen the effectiveness in the communication of public opinion, the protection of the rights of farmers' main body, the solution of problems and the realization of interests. Finally, the construction of digital countryside can improve the quality of life of farmers and boost the common prosperity of urban and rural areas. The application of digital technology realizes the online availability of public services such as education, medical care, and culture, and rural residents can satisfy their diversified and personalized needs, improve their quality of life, and narrow the gap between urban and rural public services by means of online education or e-learning. Accordingly, this paper proposes the following research hypotheses:

**Hypothesis 1**: The construction of digital villages helps to promote common wealth between urban and rural areas.

*2.2. Indirect impacts*

The construction of digital villages can drive rural industrial development and accelerate the common prosperity of urban and rural areas by improving the efficiency of agricultural production and upgrading the level of agricultural management. First of all, digital rural construction promotes the digitization of agricultural production and helps to improve agricultural production efficiency. With the continuous improvement of rural network infrastructure, as well as the development of information technology such as agricultural big data and the penetration of information technology into the field of agriculture, the digitalization of the agricultural field has been promoted, which has a positive effect on the improvement of agricultural production efficiency (Lin Haiying, 2018). At the same time, the access and use of rural Internet and other communication equipment has increased the depth and breadth of farmers' information utilization, which helps rural residents to quickly obtain effective information, such as advanced agricultural production technology, information on the supply and demand of agricultural products in the market, etc., and then make the optimal production decision, choose the appropriate varieties of planting and planting methods, and improve the efficiency of agricultural production (Li Xiaozhong, 2022). Secondly, digital village construction promotes the digitalization of agricultural production, which helps to improve the level of agricultural management. The application of big data, automation and other digital technologies in agricultural production not only promotes the intelligence and refinement of agricultural production management, but also can give birth to a group of new agricultural management subjects who know technology and management, and improve the level of agricultural management. Accordingly, this paper proposes the following research hypothesis:

**Hypothesis 2**: The construction of digital countryside can indirectly affect the common wealth of urban and rural areas by promoting the progress of agricultural technology.

The construction of digital villages can broaden the employment channels of farmers, promote the transfer of rural labor, and thus promote the realization of common prosperity in urban and rural areas. First of all, digital villages improve the level of rural information services, promote the digitization of rural life, help reduce the cost of information search for rural residents, help farmers access market information and employment opportunities, and improve the asymmetry of information in the job market. Meanwhile, the development of Internet and communication technologies in rural areas helps farmers improve their production skills through online education and online training (Aker and Ksoll, 2016). Secondly, the development of digital technology, which has changed the traditional industry model, provides a large number of opportunities for Internet-mediated entrepreneurial activities; coupled with the popularization of digital inclusive financial services in rural areas, which greatly alleviates the financing constraints of rural entrepreneurs, it will effectively promote farmers' non-farm entrepreneurial behaviors, generating a large number of jobs, in order to promote farmers' local non-farm employment. Accordingly, this paper proposes the following research hypothesis:

**Hypothesis 3**: Digital village construction can indirectly affect urban-rural common wealth by promoting rural labor transfer.

In addition, digital technology can break through resource endowment constraints and spatial and temporal limitations to a certain extent, and can produce spillover effects on other regional economic activities, which in turn affects the division of labor and cooperation between regions. Digital technology-enabled rural development promotes the spatial flow of factors such as labor, technology and capital, which helps to break the geospatial constraints, strengthens regional and urban-rural links, and realizes the sharing of knowledge and technology between regions. At the same time, digital technology has a "demonstration effect" and externality, and is easy to be learned and imitated by other regions, thus generating spatial spillover effects. It has been confirmed that the development of digital countryside has a significant spatial spillover effect on the growth of farmers' income and the improvement of agricultural total factor productivity (Sun Shuhui, 2022). Digital finance and digital economy not only affect the urban-rural income gap in the region, but also help to reduce the urban-rural income gap in neighboring regions, with certain spatial spillover effects (Yin He, 2020; Gong Lerin, 2023). Accordingly, this paper proposes the following research hypothesis:

**Hypothesis 4**: The impact of digital village construction on urban-rural common wealth has a spatial spillover effect.

## 3. Research design
### 3.1. Econometric modeling
#### 3.1.1. Benchmark model

This paper focuses on whether the construction of digital villages can promote the realization of urban-rural common wealth, and constructs the following benchmark regression model:

$$\text{CMX}_{it} = \alpha_0 + \alpha_1 \text{DIG}_{it} + \alpha_2 Z_{it} + \upsilon_{it} + \varepsilon_{it} \tag{1}$$

where i stands for province, t stands for year, $\text{CMX}_{it}$ and $\text{DIG}_{it}$ stand for urban-rural common wealth digital village construction, respectively, $Z_{it}$ are a set of control variables, $\upsilon_{it}$ time fixed effects, and $\varepsilon_{it}$ a random error term.

#### 3.1.2. Mediating effect model

In order to verify the mediating role of agricultural technology progress and rural labor force transfer in digital village construction affecting urban-rural common wealth, this paper constructs the following mediating effect model by drawing on the test of mediating effect by Wen Zhonglin (2014) and others:

$$Y_{it} = i_1 + cX_{it} + \varepsilon_{it} \tag{2}$$
$$M_{it} = i_2 + aX_{it} + \varepsilon_{it} \tag{3}$$
$$Y_{it} = i_3 + c'X_{it} + bM_{it} + \varepsilon_{it} \tag{4}$$

Where $X_{it}$ is the explanatory variable, $Y_{it}$ is the explanatory variable, $M_{it}$ is the mediating variable agricultural technology progress or rural labor force transfer. Equation (3) is the effect of core explanatory variables X on mediating variables M, and equation (4) is the effect of core explanatory variables X and mediating variables M on explained variables. If the coefficient of the independent variable a in equation (3) is significant, and at the same time the coefficients, c and c' and b of the independent and mediating variables in equations (2) and (4) are significant, it means that there is a partial mediation effect; if the coefficients of the independent variable a in equation (3) are significant and the coefficients of the independent variable c' in equation (5) are not, it means that there is a complete mediation effect.

*3.1.3. Spatial measurement models*

Spatial lag model (SLM), spatial error model (SEM) and spatial Durbin model (SDM) are three common spatial measurement models. The Spatial Durbin Model (SDM), after adding the spatial lag term between the explanatory variables and the explanatory variables, affects the explanatory variables in the region as well as in other regions, and fails to reflect the inter-subject relationship. Spatial lag model (SLM) addresses spatial dependence by adding spatial autocorrelation settings of the dependent variable, and it can test the existence of spillover effects of the dependent variable between regions. Spatial error model (SEM) reflects spatial dependence by setting the spatial autocorrelation of the error term. Therefore, this paper chooses the spatial lag model to explore the spillover effect of urban-rural common wealth between regions. The specific model is constructed as follows:

$$CMX_{it} = \beta_1 + \rho \sum_{i,j=1}^{30} W_{ij} CMX_{it} + \beta_2 DIG_{it} + \beta_3 X_{it} + \beta_4 X_{it} + \varepsilon_{it} \qquad (5)$$

Where, $\rho$ is the spatial autocorrelation coefficient, $\beta_1$、$\beta_2$、$\beta_3$、$\beta_4$ is the coefficient of each variable, $W_{ij}$ is the spatial weight matrix; $\varepsilon_{it}$ is the error term. By setting a reasonable spatial weight matrix, the spatial spillover effect can be accurately measured, which is also the fundamental work for spatial econometric analysis. Referring to the existing literature, in order to make the results more robust, the paper selects the neighbor weight matrix, geographic weight matrix and economic weight matrix for analysis.

*3.2. Variable selection and measurement*

*3.2.1. Explained variable*

common wealth between urban and rural areas (CMX). Clarifying the theoretical connotation and realistic characterization of urban-rural common wealth is the prerequisite for measuring its development level. Generally speaking, common wealth includes two dimensions: the overall "wealth" of high-quality development and the "common" wealth of sharing the fruits of development (Liu Peilin et al., 2021). This paper refers to the study of Lin and Ju Zhichao (2024), combines the availability of data at the provincial level, starts from the two dimensions of common wealth and sharing, and adds the indicators of urban and rural ecological environment in the dimension of sharing, and constructs an evaluation system with 7 secondary indicators and 11 tertiary indicators, as shown in Table 1. On this basis, the entropy weight method is used to measure, so as to get the urban-rural common wealth index of each province. The entropy weight method is an objective method of empowerment, which determines the weights of the indicators according to the influence of the relative degree of change of the indicators on the system as a whole by calculating the information entropy of the indicators, and thus the results will not be influenced by people's subjective evaluation.

**Table 1**

Comprehensive Evaluation Indicator System and Weights for Urban and Rural Common Wealth

| Primary Indicator | Secondary Indicator | Tertiary Indicator | entropy weighting |
|---|---|---|---|
| common prosperity | Urban-rural per-capita income ratio | Per capita disposable income of rural residents / per capita disposable income of urban residents | 0.046 |
| | Urban and rural per capita consumption ratio | Per capita consumption expenditure of rural residents / per capita consumption expenditure of urban residents | 0.052 |
| | The Engel coefficient ratio of urban and rural residents | Engel coefficient of urban residents / Engel coefficient of rural residents | 0.034 |
| | Integration of urban and rural industries | Added value of primary industry / added value of secondary and tertiary industries | 0.096 |
| Co-Share | The flow of urban and rural factors | Number of employees in the secondary and tertiary industries / number of employees in the primary industry | 0.430 |
| | | Urbanization rate | 0.080 |
| | Equal access to public services in both urban and rural areas | Beds in medical and health institutions per rural population / beds per urban population | 0.053 |
| | | Participation rate of urban and rural basic medical insurance | 0.051 |
| | | Per capita expenditure on education, culture and entertainment of rural residents / per capita expenditure on education, culture and entertainment of urban residents (yuan) | 0.054 |
| | Urban and rural ecological environment | forest cover | 0.089 |
| | | The harmless treatment rate of household garbage | 0.014 |

*3.2.2. Core explanatory variable*

digital village construction (DIG). Referring to the Outline of Digital Village Development Strategy and drawing on existing research results (Zhu Honggen et al., 2023), this paper focuses on the three dimensions of rural digital infrastructure, rural digital development environment and rural digital industry development (see Table 2), and utilizes entropy weighting method to measure and obtain the level of digital village construction in each region. Digital infrastructure is the foundation of digital countryside construction, which is an important guarantee to promote the digital transformation of rural industries and the hardware requirements for rural residents to enjoy information services; digital development environment provides an important guarantee to enhance the information literacy of farmers, improve the living standard of residents, and promote the development of digital industries; digital industry development is the key to the construction of digital countryside, which is an important kinetic force to realize the revitalization of rural industries and drive the income of farmers.

**Table 2**

Evaluation index system and weights of China's digital village construction level

| Primary indicator | Secondary indicator | explanation of indicator (unit) | weight |
|---|---|---|---|
| Rural digital infrastructure construction | Rural logistics construction level | Rural delivery route (km) | 0.040 |
| | Rural Internet infrastructure | Number of Internet broadband access users in rural areas (ten thousand households) | 0.093 |
| | Mobile phone penetration rate in rural areas | Mobile phone ownership per million households in rural areas (ministry) | 0.013 |
| | Computer penetration rate in rural areas | Rural per million households at the end of the computer ownership (set) | 0.032 |
| | Radio and television penetration rate in rural areas | Rural cable radio and television household registration rate (%) | 0.004 |

|  | Agrometeorological observation station | Rural meteorological observation service (piece) | 0.026 |
|---|---|---|---|
| Rural digital development environment | Rural per capita electricity consumption | Total rural electricity consumption / total rural population (kilowatt-hour / person) | 0.203 |
|  | Rural postal rate | The proportion of postal service to administrative villages (%) | 0.002 |
|  | Digital construction investment | Add value of transportation, storage and postal service (100 million yuan) | 0.052 |
|  | Level of digital services consumption | Per capita transportation and communication consumption expenditure of rural households (yuan) | 0.024 |
| Development of rural digital industry | Digital trading level | E-commerce sales volume and purchase volume (100 million yuan) | 0.138 |
|  | Rural online payment level | Rural digital financial inclusion development index | 0.038 |
|  | Digital base | All administrative villages in the Taobao village(%) | 0.334 |

Based on the collected and organized indicator data, the weights of the indicators in the comprehensive evaluation index system of urban-rural common prosperity level and digital village construction are calculated by using the entropy weight method, as shown in Tables 1 and 2, and then the indices of urban-rural common prosperity level and digital village construction level for the country as a whole and 30 provinces (autonomous regions and municipalities) are further calculated. The results show that from 2011 to 2022, the level of urban-rural common wealth and the level of digital village construction in the whole country and all provinces, municipalities and autonomous regions have increased significantly and shown accelerated development, and the level of urban-rural common wealth and the level of digital village construction have obvious regional differences, in general, the east is higher than the central and western regions, and the south is better than the north. For the limitation of space, only the national index is shown here, as in Fig. 2, from which it can be judged that there is a positive correlation between digital village construction and urban-rural common wealth.

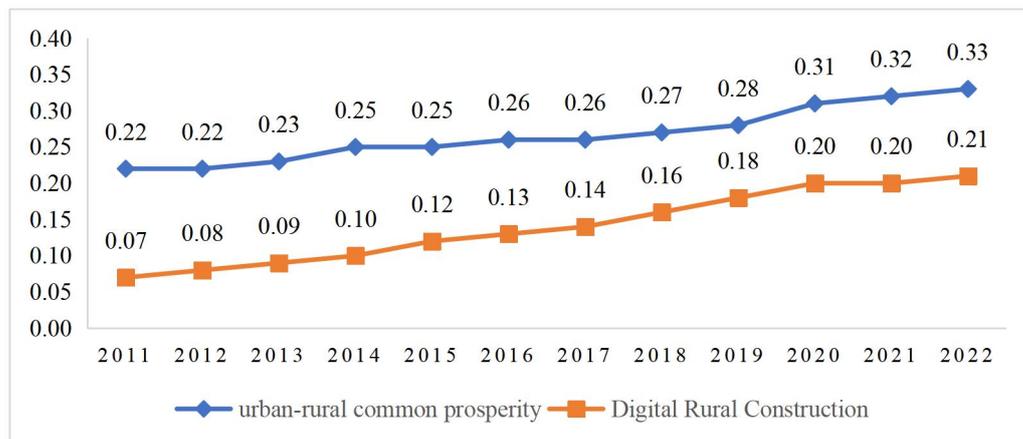

Fig. 1.National urban and rural common prosperity level and digital rural construction level index map

*3.2.3.Mediating variables*

(1) Agricultural technological progress (TFP). Referring to existing studies (Yang Yiwu and Lin Wanlong, 2016) and based on the broad agricultural technological progress, the DEA-Malmquist index method was used to measure agricultural total factor productivity to measure agricultural technological progress. The output variable is expressed as the total output value of agriculture, forestry, animal husbandry and fishery, and is deflated using the agriculture, forestry, animal husbandry and fishery index (1997=100), excluding the price factor; the input variable includes four types of inputs, namely, land input is expressed as the area sown in agriculture, labor input is measured as the number of people employed in the primary industry, machinery input is expressed as the total power of agricultural machinery, and fertilizer input is measured as the amount of agricultural Fertilizer input is measured by the amount of fertilizer applied.

(2) Rural labor force transfer (LABOR). Rural labor force transfers to non-agricultural sectors and urban areas for employment, and the direct consequence of this is that the agricultural labor force will continue to decrease. In this paper, the proportion of the number of rural workers to the rural population in each province is used to measure the scale of rural labor force transfer in each province. The measurement of the number of rural migrant workers in each province is based on the methodology of Cheng Mingwang (2018) and others, which is based on the total number of rural migrant workers in the country as well as the proportion accounted for by each province in the Rural Workers' Monitoring Survey Report for the period 2011-2022.

*3.2.4.Control variables*

(1) Regional industrial structure (STU) is measured by the share of regional tertiary industry value added in secondary industry value added.

(2) Transportation infrastructure (ROAD). Transportation infrastructure is measured by the ratio of the number of highway miles to the area of each province and region. The higher the density of roads, the more favorable the flow of factors, reducing transaction costs and increasing the level of farmers' income, thus narrowing the income gap between urban and rural areas.

(3) The level of financial support for agriculture (FIS), which is expressed as the ratio of agricultural, forestry and water conservancy expenditures in the financial expenditures of each province to the GDP of that year.

(4) The level of rural human capital (EDU), which is measured using the number of years of education per capita in rural areas. This is done by multiplying the population shares of the rural labor force in each region with different levels of education by the corresponding number of years of education. The number of years of education of the rural population is divided into five grades, i.e., 1 year, 6 years, 9 years, 12 years and 16 years, corresponding to illiterate and semi-illiterate, elementary school, junior high school, senior high school, and college and above, respectively.

(5) Openness to the outside world (OPEN). It is expressed as the ratio of import and export volume of each province to the Gross Regional Product, and the import and export volume of each province has been converted with the exchange rate of the current year.

*3.3. Data sources*

The data in this paper mainly come from China Statistical Yearbook, China Rural Statistical Yearbook, China Population and Employment Statistical Yearbook, provincial statistical yearbooks and statistics, and Digital Inclusive Finance Index of Peking University. The panel data of 30 provinces (autonomous regions and municipalities) in mainland China from 2011 to 2022 are finally selected, and all variables are calculated and organized by the authors. The statistical description of each variable is shown in Table 3.

**Table 3**
Descriptive statistics of variables

| variable | N | mean value | standard deviation | minimum value | maximal value |
| --- | --- | --- | --- | --- | --- |
| urban-rural common prosperity (CMX) | 360 | 0.266 | 0.065 | 0.124 | 0.619 |
| Digital rural construction level (DIG) | 360 | 0.141 | 0.105 | 0.018 | 0.603 |
| Agricultural technology progress (TDP) | 360 | 1.187 | 0.108 | 1.022 | 1.628 |
| Scale of rural labor force transfer | 360 | 0.425 | 0.192 | 0.063 | 0.940 |
| (LABOR) | | | | | |
| Regional industrial structure (STU) | 360 | 1.374 | 0.870 | 0.578 | 5.858 |
| Level of transportation infrastructure (ROAD) | 360 | 0.989 | 0.549 | 0.089 | 2.310 |
| Financial support for agriculture (FIS) | 360 | 0.117 | 0.036 | 0.041 | 0.288 |
| Rural human capital level (EDU) | 360 | 7.903 | 0.585 | 6.119 | 10.071 |
| Open to the outside world (OPEN) | 360 | 0.265 | 0.285 | 0.002 | 1.464 |

## 4. Analysis of empirical results

*4.1 Baseline regression*

Table 4 reports the regression results of the impact of digital village construction on urban-rural shared prosperity, where columns (1), (2), and (3) show the results of parameter estimation using least squares, fixed effects model, and systematic GMM methods, respectively. In Column (3), AR (1) passes the significance test, indicating the existence of first-order autocorrelation in the differenced disturbed term, while AR (2) fails the significance test, indicating that the original hypothesis that there is no second-order autocorrelation in the differenced disturbed term could not be rejected, i.e., the model does not suffer from autocorrelation problem and overcomes the endogeneity problem better. Also from the Hansen test results, the model does not have over-identification problem. The first-order lagged term of the explanatory variable, i.e., urban-rural common wealth, is significantly positive at the 1% level, which also indicates that the development of urban-rural common wealth has continuity, and the selection of the dynamic panel model for analysis is reasonable. From the results of columns (1), (2) and (3) in Table 4, the estimated coefficients of digital village construction are 0.156, 0.241 and 0.042 respectively, and they pass the significance test of 1%, indicating that the digital village construction can significantly promote the urban-rural common wealth. The reason is that, on the one hand, the digital countryside improves the level of rural information services, which helps to alleviate the "information poverty" in rural areas, and at the same time, the digital countryside promotes the non-farming entrepreneurship of farmers, as well as the new forms of business and new products spawned by digital technology, which is conducive to the increase of jobs, and thus promotes the growth of farmers' incomes, and continuously reduces the income disparity between urban and rural areas. The gap between urban and rural incomes is constantly narrowing. On the other hand, the integration of digital technologies such as big data and the Internet of Things (IoT) with agricultural production has promoted the development of agricultural intelligence and refinement, which has helped to improve the level of agricultural technology and the efficiency of agricultural production, and improve the efficiency of resource allocation, thus driving agricultural economic growth. Research hypothesis 1 was verified. Among the control variables, upgrading the regional industrial structure, i.e. increasing the proportion of the tertiary industry, strengthening the construction of rural transportation infrastructure, increasing financial support for agriculture, and improving the level of opening up to the outside world all help to promote the realization of urban-rural common wealth.

**Table 4**
Benchmark regression results

| variable | (1) | (2) | (3) |
|---|---|---|---|
| | OLS | FE | SYS-GMM |
| DIG | 0.156*** | 0.241*** | 0.042*** |
| | (0.029) | (0.031) | (0.005) |
| L.CMX | | | 1.054*** |
| | | | (0.014) |
| STU | 0.006*** | 0.001 | 0.001*** |
| | (0.002) | (0.001) | (0.000) |
| ROAD | 0.027*** | 0.043*** | 0.004** |
| | (0.006) | (0.014) | (0.002) |
| FIS | 0.329*** | 0.394*** | 0.052** |
| | (0.107) | (0.092) | (0.025) |
| EDU | 0.037*** | 0.060*** | -0.016*** |
| | (0.005) | (0.009) | (0.001) |
| OPEN | 0.034** | -0.052*** | 0.007*** |
| | (0.013) | (0.017) | (0.002) |
| AR(1) | | | 0.011 |
| AR(2) | | | 0.241 |
| Hansen Test | | | 1.000 |
| _cons | -0.134*** | -0.320*** | 0.101*** |
| | (0.046) | (0.068) | (0.014) |
| N | 360 | 360 | 330 |
| R2 | 0.430 | 0.579 | |

Note: Standard error is in parentheses, "*, * * and * * *" indicate the significant levels of 10%, 5% and 1% respectively, the same as in the table below.

### 4.2. Robustness tests

In order to ensure the robustness of the study's conclusions, the following three aspects are further tested for robustness: First, the sample of municipalities directly under the central government is excluded. Given that the four municipalities of Beijing, Tianjin, Shanghai and Chongqing have more favorable factor endowment conditions and policy advantages, the sample of the four municipalities is excluded and regressed again. The second is to replace the explanatory variables. Referring to the existing literature, the income gap between urban and rural residents is used to measure the common wealth of urban and rural areas, and the income gap between urban and rural residents is expressed by the Thiel index. Third, considering the possible lag in the impact of digital village construction on urban-rural common wealth, the lagged one-period term for the level of digital village construction is introduced and re-estimated using the two-stage least squares method to solve the possible endogeneity problem. The results show (Table 5) that digital village construction still has a significant promotion effect on urban-rural common wealth, further validating the findings of this paper.

**Table 5**
Robustness test

| variable | (1) Excluding municipalities | (2) Substitution of explanatory variables | (3) Instrumental variable method (2SLS) |
|---|---|---|---|
| DIG | 0.256*** | -0.129*** | 0.158*** |
| | (0.026) | (0.019) | (0.030) |
| control variable | YES | YES | YES |
| Constant term | -0.289*** | 0.560*** | -0.117** |
| | (0.062) | (0.042) | (.047) |
| Unrecognized Test Anderson LM statistic | | | 316.119*** |

|  | Weak Instrumental Variables Cragg-Donald Wald F statistic |  |  | 7355.894*** |
|---|---|---|---|---|
|  | N | 312 | 360 | 330 |

### 4.3. Heterogeneity analysis
#### 4.3.1. Regional heterogeneity

Due to the vastness of China, there are large differences in resource endowments among regions, leading to large differences in the level of economic development. Therefore, considering the uneven development of China's regions, the country is divided into eastern and central-western regions to examine the regional differences in the impact of digital village construction on urban-rural common wealth. The estimation results are shown in columns (1), (2) and (3) of Table 6. The results show that digital rural construction passes the significance test in both eastern and central-western regions, but in terms of the effect, the regression coefficients of digital rural construction in central-western provinces are larger than those in eastern provinces, which indicates that the role of digital rural construction in promoting urban-rural common wealth is more prominent in central-western regions. The possible reasons are: the real endowment of the central and western regions is quite different from that of the eastern regions, whether it is infrastructure, monetary capital, human resources, or technology level are lagging behind that of the eastern regions, and digital rural construction provides a new window of opportunity for the development of rural areas in the central and western provinces, which can make up for the multiple disadvantages of the late-developed regions, and empower urban and rural common prosperity.

#### 4.3.2. Panel quantile regression

The previous analysis shows that digital village construction can significantly promote urban-rural common wealth, but this is only the average degree of impact between variables, and cannot examine the differences in the distribution of urban-rural common wealth development levels. For this reason, a panel quantile regression model is further used to explore the heterogeneous impact of digital village construction on the development level of urban-rural common wealth. In this paper, five representative quantile points 0.10, 0.25, 0.50, 0.75 and 1.90 are set, and the estimation results are shown in Table 6, Column (4), Column (5), Column (6), Column (7) and Column (8). The results show that digital village construction has a positive contribution to urban-rural common wealth at different quartiles, except that it is not statistically significant at the Q=0.10 and Q=0.50 quartiles. In terms of the size of the coefficients, digital village construction has a greater and more significant contribution to the provinces in the high quartile of urban-rural common wealth level. The reason may lie in the fact that provinces with a higher level of development of common wealth between urban and rural areas have a better rural digital infrastructure, while rural residents are more culturally literate, have a better ability to master and utilize digital technology, and have a higher fit between digital technology and agricultural production, so they are able to share the digital dividend.

**Table 6**
Results of heterogeneity estimation

| variable | (1) | (2) | (3) | (4) | (5) | (6) | (7) | (8) |
|---|---|---|---|---|---|---|---|---|
|  | Eastern | Central | Western | Q=0.10 | Q=0.25 | Q=0.50 | Q=0.75 | Q=0.90 |
| DIG | 0.184*** | 0.659*** | 0.857*** | 0.088 | 0.231* | 0.261 | 0.311*** | 0.381** |
|  | (0.050) | (0.116) | (0.091) | (0.086) | (0.136) | (0.160) | (0.100) | (0.184) |
| STU | 0.001 | 0.043*** | -0.003 | 0.000 | 0.003* | 0.001 | -0.000 | 0.002 |
|  | (0.002) | (0.011) | (0.010) | (0.001) | (0.002) | (0.004) | (0.002) | (0.003) |
| ROAD | 0.006 | -0.024 | 0.012 | -0.031 | 0.048 | 0.131** | 0.135** | -0.044 |
| | (0.023) | (0.023) | (0.019) | (0.044) | (0.041) | (0.057) | (0.065) | (0.079) |
| FIS | 0.309 | 0.012 | 0.165** | 1.047* | 0.228 | 0.331* | 0.384 | 0.017 |
| | (0.414) | (0.104) | (0.079) | (0.551) | (0.303) | (0.180) | (0.297) | (0.295) |
| EDU | 0.050*** | -0.003 | 0.048*** | 0.114** | 0.039 | 0.028 | 0.007 | 0.057 |
| | (0.016) | (0.012) | (0.012) | (0.043) | (0.029) | (0.027) | (0.031) | (0.046) |
| OPEN | -0.092*** | -0.066 | 0.030 | 0.055 | 0.029 | 0.058 | 0.043 | -0.148 |
| | (0.025) | (0.074) | (0.049) | (0.047) | (0.045) | (0.047) | (0.047) | (0.110) |
| _cons | -0.143 | 0.210** | -0.234*** | -0.827** | -0.199 | -0.186 | 0.020 | -0.084 |
| | (0.139) | (0.097) | (0.084) | (0.350) | (0.222) | (0.210) | (0.231) | (0.352) |
| N | 132 | 96 | 132 | 360 | 360 | 360 | 360 | 360 |

*4.4. Mechanism of action tests*

The results of the previous study show that digital village construction can significantly contribute to the realization of urban-rural common wealth. The next step is to further test the internal mechanism of digital village construction affecting urban-rural common wealth. Estimation is carried out according to the mediation effect model constructed in the previous section. The test results are shown in Table 7.

The regression results in Column (2) show that digital village construction can significantly promote agricultural technological progress. This is due to the fact that the development of agricultural informatization accelerates the flow of information, which helps farmers to obtain advanced agricultural production technology, information on supply and demand in the agricultural market, etc., so as to make optimal production decisions and improve the efficiency of resource allocation. At the same time, the integration of big data, the Internet of Things and other digital technologies with agricultural production has promoted the development of agricultural intelligence and refinement, and the transformation of the agricultural production model will prompt agricultural development to shift from factor-driven to technology-driven, which will help the dissemination and diffusion of advanced technologies, and thus promote the technological progress of agricultural technology.

The results in column (3) show that the construction of digital countryside has a significant role in promoting the transfer of rural labor. The reason is that, on the one hand, the digital village promotes the digitization of rural life, which is conducive to reducing the information search cost of rural residents, improving the information asymmetry of the job market, and giving rural residents more opportunities to enter the non-agricultural sector for employment. On the other hand, the development of digital technology has changed the traditional industry model, providing a large number of opportunities for Internet-mediated entrepreneurial activities; coupled with the popularization of digital inclusive financial services in rural areas, which greatly alleviates the financing constraints of rural entrepreneurs, it will effectively promote farmers' non-farm entrepreneurial behavior and promote local non-farm employment of agricultural labor.

The results in Column (5) show that the estimated coefficients of both digital rural construction and agricultural technological progress are significantly positive, indicating the existence of the mediating effect of agricultural technological progress, i.e., digital rural construction can realize urban-rural common wealth by promoting agricultural technological progress. Column (6) shows that the estimated coefficients of digital village construction and rural labor force transfer are both significantly positive and pass the 1% significance test, which means that digital village construction can realize urban-rural common wealth by promoting rural labor force transfer.

**Table 7**
Estimated results of intermediation effects

| variable | (1) urban-rural common prosperity | (2) Agricultural technology progress | (3) rural labor migration | (4) urban-rural common prosperity | (5) urban-rural common prosperity |
|---|---|---|---|---|---|
| DIG | 0.241*** | 0.298*** | 0.375*** | 0.146*** | 0.216*** |
|  | (0.031) | (0.044) | (0.054) | (0.029) | (0.033) |
| TFP |  |  |  | 0.317*** |  |
|  |  |  |  | (0.034) |  |
| LABOR |  |  |  |  | 0.067** |
|  |  |  |  |  | (0.031) |
| STU | 0.001 | 0.001 | 0.002 | 0.001 | 0.001 |
|  | (0.001) | (0.002) | (0.002) | (0.001) | (0.001) |
| ROAD | 0.043*** | 0.078*** | 0.165*** | 0.019 | 0.032** |
|  | (0.014) | (0.020) | (0.024) | (0.012) | (0.015) |
| FIS | 0.394*** | 0.660*** | -0.014 | 0.186** | 0.395*** |
|  | (0.092) | (0.133) | (0.163) | (0.084) | (0.091) |
| EDU | 0.060*** | 0.121*** | 0.047*** | 0.022** | 0.057*** |
|  | (0.009) | (0.013) | (0.016) | (0.009) | (0.009) |
| OPEN | -0.052*** | 0.067*** | 0.151*** | -0.073*** | -0.062*** |
|  | (0.017) | (0.025) | (0.030) | (0.015) | (0.017) |
| _cons | -0.320*** | 0.012 | -0.204* | -0.324*** | -0.307*** |
|  | (0.068) | (0.098) | (0.120) | (0.060) | (0.068) |
| N | 360 | 360 | 360 | 360 | 360 |

*4.5. Spatial spillover effects*

*4.5.1. Spatial correlation analysis*

Before spatial measurement regression, spatial correlation test is needed for key core variables. Therefore, this paper calculates the global Moran index of urban-rural common wealth and digital village construction level of Chinese provinces during 2011-2022, and the results are shown in Table 8. It can be seen that the levels of urban-rural common wealth and digital village construction in China's provinces have positive spatial correlation and exhibit spatial clustering characteristics, and all of them have passed the significance test. Therefore, it is reasonable and necessary to use spatial econometric models to analyze their spatial spillover effects.

**Table 8**
Spatial autocorrelation test

| YEAR | urban-rural common prosperity | | | YEAR | Digital rural construction | | |
|---|---|---|---|---|---|---|---|
|  | Moran'I | Z-value | P-value |  | Moran'I | Z-value | P-value |
| 2011 | 0.442 | 3.754 | 0.000 | 2011 | 0.372 | 3.258 | 0.001 |
| 2012 | 0.489 | 4.126 | 0.000 | 2012 | 0.371 | 3.270 | 0.001 |
| 2013 | 0.459 | 3.892 | 0.000 | 2013 | 0.469 | 4.148 | 0.000 |
| 2014 | 0.506 | 4.259 | 0.000 | 2014 | 0.459 | 4.092 | 0.000 |
| 2015 | 0.527 | 4.415 | 0.000 | 2015 | 0.488 | 4.214 | 0.000 |
| 2016 | 0.510 | 4.285 | 0.000 | 2016 | 0.487 | 4.208 | 0.003 |
| 2017 | 0.471 | 3.989 | 0.000 | 2017 | 0.472 | 4.081 | 0.000 |

| 2018 | 0.447 | 3.813 | 0.000 | 2018 | 0.468 | 4.100 | 0.000 |
| 2019 | 0.413 | 3.540 | 0.000 | 2019 | 0.443 | 3.968 | 0.000 |
| 2020 | 0.303 | 2.778 | 0.005 | 2020 | 0.480 | 4.218 | 0.000 |
| 2021 | 0.284 | 2.657 | 0.007 | 2021 | 0.355 | 3.240 | 0.001 |
| 2022 | 0.259 | 2.557 | 0.010 | 2022 | 0.358 | 3.265 | 0.001 |

The Moran scatterplot was further adopted to reflect the spatial agglomeration of urban-rural common wealth in each provincial area, and the four quadrants of the Moran scatterplot showed the spatial characteristics of high-high agglomeration, bottom-high agglomeration, low-low agglomeration, and high-low agglomeration in that order. The results are shown in Figure 1. It can be seen that the development level of urban-rural common wealth in each province is uneven in spatial distribution, and shows significant "High-High (H-H)" and "Low-Low (L-L)" spatial agglomeration characteristics among provinces. The first quadrant of high-high (H-H) agglomeration mainly includes Jiangsu, Zhejiang, Hainan and other eastern coastal provinces, while the third quadrant of low-low (L-L) agglomeration mainly includes Shanxi, Qinghai, Inner Mongolia and other central and western provinces. This reflects the higher level of development of urban-rural shared prosperity in the more economically developed eastern coastal provinces and the lower level of development of urban-rural shared prosperity in the less economically developed central and western provinces.

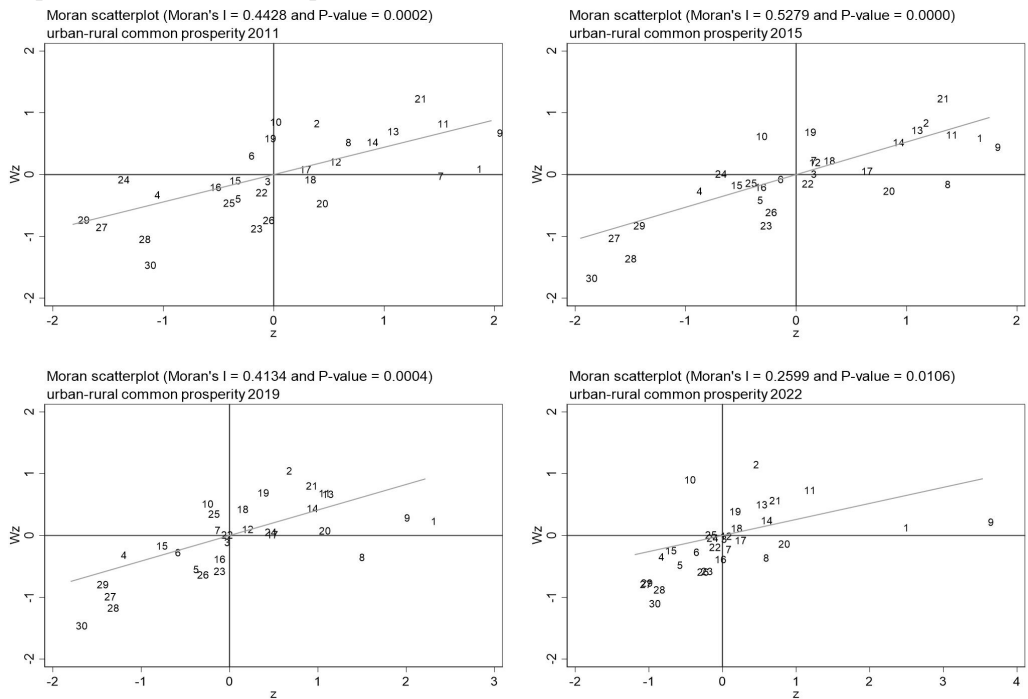

**Fig. 2.** Local Moran scatter plot of the urban-rural common wealth index in 2012,2015,2019 and 2022 4.5.2.Spatial econometric model regression results

Table 8 reports the estimation results of the spatial lag model (SLM) under the three spatial weight matrices. The results show that the spatial autoregressive coefficients are positive and pass the 1% significance test under the three spatial weight matrices, indicating that there is a spatial dependence between urban-rural common wealth in each province of China, and that an increase in the level of urban-rural common wealth in the local region will promote urban-rural common wealth in other regions. The estimated coefficients of digital village construction on urban-rural common wealth are all significantly positive, indicating that digital village construction will promote urban-rural common wealth in other regions by promoting local urban-rural common wealth.

The effect decomposition is further carried out through the partial differential method to calculate the direct effect, indirect effect and total effect of digital village construction. The results show that under the three spatial weight matrices, the direct, indirect and total effects of digital rural construction are significantly positive, which indicates that digital rural construction is not only conducive to the realization of urban-rural common wealth in the local province, but also helps to promote the realization of common wealth in the neighboring provinces, which has a significant spatial spillover effect. This is due to the fact that digital technology-enabled rural development promotes the spatial mobility of factors such as labor, technology and capital, and is conducive to the sharing of knowledge and agricultural technology among regions; at the same time, digital technology has a "demonstration effect" and externality, and is easy to be learned and imitated by other regions, thus generating a spatial spillover effect.

**Table 8**
Estimated results of the spatial measurement model

| variable | (1) Adjacent weight matrix | (2) Geographic weight matrix | (3) Economic weight matrix |
| --- | --- | --- | --- |
| DIG | 0.061** | 0.047* | 0.051** |
|  | (0.025) | (0.025) | (0.021) |
| direct effect | 0.073** | 0.049* | 0.062** |
|  | (0.030) | (0.026) | (0.025) |
| indirect effect | 0.085** | 0.052* | 0.147** |
|  | (0.033) | (0.027) | (0.059) |
| Total effect | 0.158** | 0.101* | 0.209** |
|  | (0.062) | (0.053) | (0.083) |
| control variable | control | control | control |
| ρ | 0.613*** | 0.537*** | 0.755*** |
|  | (0.041) | (0.033) | (0.033) |
| N | 360 | 360 | 360 |

### 5. Conclusions and insights

Based on the provincial panel data from 2011 to 2022, this paper mainly studies the effect of digital rural construction on urban-rural common wealth and its mechanism. The results show that, firstly, overall, digital rural construction can significantly promote the realization of urban-rural common wealth, and regionally, the impact effect of digital rural construction on urban-rural common wealth is characterized by "high in the central and western regions and low in the eastern regions". Secondly, digital rural construction mainly drives urban-rural common wealth through two paths: enhancing agricultural technology progress and promoting rural labor transfer. Thirdly, there is an obvious spatial spillover effect of digital rural construction for urban-rural common prosperity at the regional level, i.e., digital rural construction can not only effectively promote urban-rural common prosperity in the region, but also have a positive effect on the development of urban-rural common prosperity in neighboring regions.

Based on the above findings, the following policy insights have been obtained: first, it is necessary to improve the level of agricultural digital development, innovate and develop smart agriculture, and drive agricultural economic growth and increase farmers' income through agricultural digital transformation, so as to realize urban-rural common wealth. On the one hand, it is necessary to strengthen the construction of rural digital infrastructure, develop a new generation of information networks, promote universal mobile network services, and use big data to build information databases on agricultural production and sales of agricultural products, so as to provide a solid foundation for the digitization and intelligentization of agricultural production. On the other hand, we will promote the deep integration of "Internet+Agriculture", make use of the Internet, big data and other information technologies to promote and apply them in agricultural production, and accelerate the development of digital agriculture, such as the construction of intelligent greenhouses and intelligent irrigation, the development of digital farming, and the promotion of automation of agricultural machinery and equipment, so as to realize the refinement and automation of agricultural production. Secondly, it is necessary to guide the integration of digital technology and traditional agriculture, promote the integration of digital technology and primary, secondary and tertiary industries in rural areas, extend the industrial chain, increase the number of jobs for farmers, and actively guide rural residents to transfer employment locally to provide a good help to farmers to increase their income, in order to gradually narrow the income gap between urban and rural residents. Finally, it is necessary to strengthen inter-regional exchanges and cooperation, actively break through the "local" mentality, optimize the spatial layout of digital rural development, and give full play to the role of radiation and impetus of the advanced regions in digital rural development to the backward regions.

**Acknowledgments**

This research was supported by the General Program of Hubei Provincial Social Science Foundation (HBSK2022YB347),and the Philosophy and Social Science Youth Program of Hubei Provincial Department of Education(22Q180).